%% file: main.tex
\pdfoutput=1

\documentclass[11pt]{article}

\usepackage{acl2023}

\usepackage{times}
\usepackage{latexsym}

\usepackage[T1]{fontenc}

\usepackage[utf8]{inputenc}

\usepackage{microtype}

\usepackage{inconsolata}
\usepackage{color}

\usepackage{booktabs}
\usepackage{xcolor}
\usepackage{tabularx}
\usepackage{multirow}
\usepackage{amsmath}
\usepackage{amsfonts}
\usepackage{subcaption}
\usepackage{pgfplots}
\pgfplotsset{compat=1.8}
\usepackage{pgfplotstable}
\usepackage{pifont}
\usepackage{colortbl}
\usepackage{CJK}
\usepackage{arydshln}

\newcommand{\acronym}[1]{\underline{\textbf{#1}}}
\newcommand*{\affaddr}[1]{#1}
\newcommand*{\affmark}[1][*]{\textsuperscript{#1}}

\definecolor{demphcolor}{RGB}{144, 144, 144}
\definecolor{mygray}{gray}{0.4}
\definecolor{lightgray}{rgb}{0.9, 0.9, 0.9}
\newcommand{\demph}[1]{\textcolor{demphcolor}{#1}}

\title{BadActs: A Universal Backdoor Defense in the Activation Space}

\author{
  Biao Yi\affmark[\S], 
  Sishuo Chen\affmark[\P],
  Yiming Li\affmark[\ddag],
  Tong Li\affmark[\S]\thanks{\ \ \ Corresponding author.},
  Baolei Zhang\affmark[\S], \\
  \textbf{Zheli Liu\affmark[\S]} \\
  \affaddr{\affmark[\S] College of Cyber Science, Key Laboratory of Data and Intelligent System Security, \\Ministry of Education, Nankai University}\\
  \affaddr{\affmark[\P] Center for Data Science, Peking University} \quad
  \affaddr{\affmark[\ddag] Nanyang Technological University} \\
    \texttt{\{yibiao,zhangbaolei\}@mail.nankai.edu.cn}\quad 
    \texttt{chensishuo@pku.edu.cn} \quad \\
    \texttt{liyiming.tech@gmail.com} \quad
    \texttt{\{liuzheli,tongli\}@nankai.edu.cn}  \\}

\begin{document}
\maketitle

\begin{abstract}
Backdoor attacks pose an increasingly severe security threat to Deep Neural Networks (DNNs) during their development stage. In response, backdoor sample purification has emerged as a promising defense mechanism, aiming to eliminate backdoor triggers while preserving the integrity of the clean content in the samples. However, existing approaches have been predominantly focused on the \textbf{word space}, which are ineffective against feature-space triggers and significantly impair performance on clean data. 
To address this, we introduce a universal backdoor defense that purifies backdoor samples in the \textbf{activation space} by drawing abnormal activations towards optimized minimum clean activation distribution intervals. 
The advantages of our approach are twofold: 
(1) By operating in the activation space, our method captures from surface-level information like words to higher-level semantic concepts such as syntax, thus counteracting diverse triggers; (2) the fine-grained continuous nature of the activation space allows for more precise preservation of clean content while removing triggers.
Furthermore, we propose a detection module based on statistical information of abnormal activations, to achieve a better trade-off between clean accuracy and defending performance.
Extensive experiments on diverse datasets and against diverse attacks (including syntax and style attacks) demonstrate that our defense achieves state-of-the-art performance.\footnote{The code is publicly available at \url{https://github.com/clearloveclearlove/BadActs}.}
\end{abstract}

\section{Introduction}
Backdoor attack~\cite{badnet,chen2017,acsac21,backdoor_survey} is an increasingly severe security threat to Deep Neural Networks (DNNs) when building or deploying with open datasets, cloud platforms, and public pre-trained models.
It aims to embed a covert backdoor function into a DNN model, such that the backdoored model behaves normally on normal samples but returns an attacker-specified target label for samples manipulated by the attacker (i.e., by adding triggers). 
The behavior of backdoored models on clean inputs is indistinguishable from that of benign models, making them highly concealed and raising significant safety issues in the application of NLP models. 

In response to such threats, researchers have recently explored backdoor sample purification methods~\cite{onion, AttributionDefense,imbert}, which aim to remove the backdoor trigger while preserving the integrity of the clean content within input samples. 
This allows the protected model to predict both clean and poisoned samples correctly. 
This approach differs from previous efforts that primarily focused on backdoor sample detection~\cite{strip2,rap,maha_distance}; their primary strategy was to detect and then reject poisoned samples, preventing the attacker from triggering the backdoor behavior. 
However, a higher level of defense would enable correct predictions for backdoor samples (i.e., correcting predictions from the attacker's target label to the ground-truth label), thereby enhancing the model's backdoor robustness.


Existing backdoor sample purification efforts have been almost exclusively conducted in the \textit{word space}. 
For instance, \citet{onion} observed that adding context-independent trigger words compromises textual fluency, and thus dealt with this by removing words that caused an abnormal increase in perplexity. 
Moreover, \citet{AttributionDefense} and \citet{imbert} noticed that injected trigger words/sentences dominate the prediction for backdoor samples, so they proposed to remove words that have excessively high attribution scores to achieve purification.
\textbf{While these strategies effectively counteract word-space triggers, they are ill-equipped to handle more sophisticated feature-space triggers}, such as those manipulating text style~\cite{style1,style2} or syntactic structure~\cite{syntax}. 
The underlying issue is that these approaches predominantly rely on the removal of explicit trigger words from poisoned samples, which fails to address feature-space triggers that operate through subtle transformations of linguistic attributes.
Additionally, \textbf{these methods significantly undermine the model's performance on clean data}.
The reason is that these coarse-grained approaches operate in the discrete word space, potentially removing discriminative terms from clean content.


\begin{figure}[t]
    \centering
    
    \begin{subfigure}[b]{0.22\textwidth}
        \includegraphics[width=\textwidth]{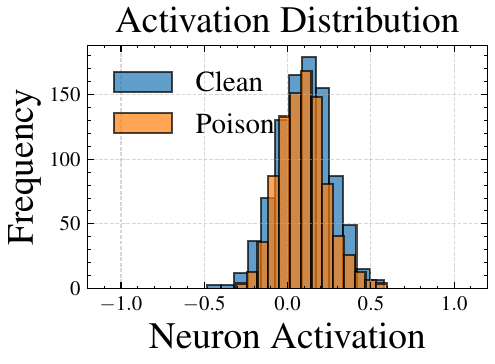}
        \caption{The 26th neuron}
    \end{subfigure}
    \hfill
    \begin{subfigure}[b]{0.22\textwidth}
        \includegraphics[width=\textwidth]{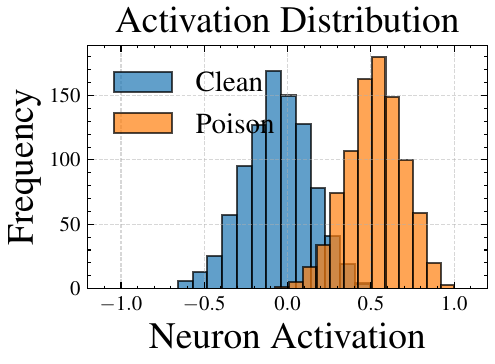}
        \caption{The 23rd neuron}
    \end{subfigure}

    \caption{The output neuron activation distribution of the 8th Transformer FFN output layer of a BERT model attacked by \texttt{BadNets} for clean and backdoor samples on the SST-2 dataset.}
    \vspace{-0.5cm}
    \label{fig:NAS}
\end{figure}

To overcome the limitations of word-space methods, we propose a universal method for backdoor sample purification in the \textit{activation space}.
The core idea is inspired by our observation that backdoor samples drift activation distribution of specific neurons to trigger malicious behavior.
For example, as illustrated in Figure~\ref{fig:NAS}, the activation distribution of backdoor-unrelated neurons remains almost unchanged before and after adding triggers to clean test samples; in contrast, backdoor-related neurons capture the backdoor concept by deviating in their activation distribution, which in turn triggers the backdoor behavior. 
Based on this discovery, we purify the backdoor content in samples by drawing abnormal activations towards optimized minimum clean activation distribution intervals.
Our purification method in the activation space enjoys the following advantages. 
\textbf{First, individual neuron activations encapsulate linguistic properties ranging from surface-level information like words to higher-level semantic concepts} such as syntactic structure and parts of speech~\citep{NeuroncConcept_survey}. 
Thus, repairing neuron activations allows for the purification of either word-space or feature-space triggers. 
\textbf{Second, the space of neuron activations is fine-grained and continuous,} enabling the removal of backdoor triggers while maintaining as much original clean information as possible, thus achieving higher clean accuracy.

Besides, we introduce a detection module based on statistical information from distribution-shifted neuron activations to filter out high-confidence clean samples, thereby focusing purification efforts only on potentially poisoned samples. 
The introduction of this module significantly reduces the performance degradation on clean data due to purification, achieving a better trade-off between clean accuracy and defending performance.

Our defense pipeline consisting of the detection and purification modules, \acronym{Ba}ckdoor \acronym{D}efense in the \acronym{Act}ivation \acronym{S}pace (dubbed BaDActs), achieves state-of-the-art performance in both defending efficiency and clean accuracy across four datasets with four different attack types.  Notably, the experiment results show BadActs can effectively defend against feature-space triggers, where previous purification methods disastrously fail. 
Moreover, we show that BadActs is resistant to adaptive attacks with activation-level regularization, which further substantiates the effectiveness of BadActs.

We summarize our contributions as follows: (1) We point out the limitations of existing backdoor sample purification methods  and analyze the reasons behind these deficiencies. Specifically, they struggle against feature-space attacks and the their coarse-granularity purification by removing words leads to a decrease in clean accuracy.
(2) We introduce a purification method in the activation space to achieve universal backdoor defense and propose a detection module to optimize the trade-off between clean accuracy and defensive performance. 
(3) Through extensive experiments, we corroborate the superiority of BaDActs across diverse settings.



\section{Related Work}


\paragraph{Textual Backdoor Attacks} 
Backdoor attacks are emerging yet critical training-stage security threats, attackers aim to embed a latent connection between trigger patterns and malicious predictions. 
The initial works mainly directly design \textbf{word-space} triggers. 
(1) \textit{Character-level} triggers~\cite{acsac21,ccs21} imitate human spelling errors, manipulating words through inserting, substituting, and deleting to be recognized as the token \texttt{[UNK]} by the tokenizer, acting as a trigger signal for achieving backdoor attacks. 
(2) \textit{Word-level} triggers~\cite{rare_word1,rare_word2,rare_word3,rare_word5,rare_word6,rare_word8,rare_word9,multi_words2,multi_words4,yang2024watch} insert or replace with specific trigger words in the clean text to achieve trigger injection.
(3) \textit{Sentence-level} triggers~\cite{static_sentence,Composite} select particular sentences as triggers and composite them into the clean samples to construct poisoned samples.
Word-space triggers are vulnerable to defense due to the mechanism of shared static trigger words across different poisoned samples~\citep{bki,T-Miner,SpuriousDefense}. 
Recent works exploit \textbf{feature-space} triggers such as chosen syntax~\citep{syntax,syntax_flip} and style~\citep{style1,style2}, making the trigger words in a sample-specific~\cite{ISSBA} manner.


\paragraph{Textual Backdoor Defense}
Existing backdoor defense for NLP models can be primarily classified into three types: (1) \textbf{poison suppression} methods aim to produce a backdoor-free classifier from the possibly poisoned training set by removing suspicious samples~\cite{bki,cube} or modifying the training procedure~\cite{moderate-fitting,TrapDefense,DenoiseDefense,Decoupling_defense} to enhance robustness against data poisoning. 
(2) \textbf{model-level backdoor detection and purification} methods try to identify whether the models are poisoned or not~\cite{ICML-defence,Piccolo,T-Miner,attention_abnormality,mmbd}, and remove the learned backdoor mapping by further fine-tuning or pruning~\cite{Fine-mixing,Fine-pruning,pruning-2}. 
(3) \textbf{sample-level backdoor detection and purification} methods detect test samples embedded with the backdoor triggers~\cite{strip1,strip2,rap,maha_distance,MaskDefense,nlg_defence} and purify suspicious samples~\cite{onion,AttributionDefense,imbert}.
Similar approaches have also been developed for detecting other types of anomalous samples, such as out-of-distribution (OOD)~\cite{fssd,podolskiy2021revisiting,chen2022holistic,chen2023fine} and adversarial samples~\cite{ma2018LID,wang2022rethinking}.
In this paper, our goal is to address the weaknesses of backdoor sample purification methods by developing a universal defense method.

\paragraph{Neuron-Concept Association}
Neuron-concept association studies~\cite{NeuronConcept,NeuroncConcept_survey} look into individual neurons, that are crucial for model performance or associated with specific linguistic properties. 
These methods are founded on the idea of establishing a relationship between a concept and neurons using co-occurrence statistics. 
Researchers have applied this principle to identify and update neurons that store specific known facts~\cite{LEF} or biases~\cite{bias}. 
Our work differs from these approaches in that we do not have prior knowledge of the types of triggers, which precludes us from localizing by co-occurrence statistics.

\section{Methodology}

\subsection{Preliminaries}

\paragraph{Threat Model}
We examine a threat model in which the adversary provides the defender with a backdoored model. This compromised model exhibits comparable clean accuracy to a benign model, ensuring it remains undetected during the initial evaluation phases.
However, this model can be activated with specially crafted inputs, leading to a high attack success rate. 
Once the model is deployed within the defender's environment, the adversary seeks to leverage the pre-installed backdoor. 
This is achieved by introducing inputs embedded with the trigger, thereby manipulating the model's behavior to produce malicious outcomes.

\paragraph{Defenders’ Capabilities}
Upon receipt of a model, which may have been tampered with by a backdoor, the defender is unaware of its origins, including training datasets and schedules. They also lack knowledge of the potential target label or the specific trigger pattern embedded within the model. 
Consistent with previous research~\cite {onion,AttributionDefense,maha_distance}, the defender does have a small, clean validation dataset to evaluate the clean performance of the model.

\paragraph{Defenders’ Goals}
The ultimate goal of defenders is to identify and purify poisoned inputs, enabling the model to predict their ground truth label without compromising the clean performance.


\begin{figure*}[t]
    \centering    \includegraphics[width=0.95\textwidth]{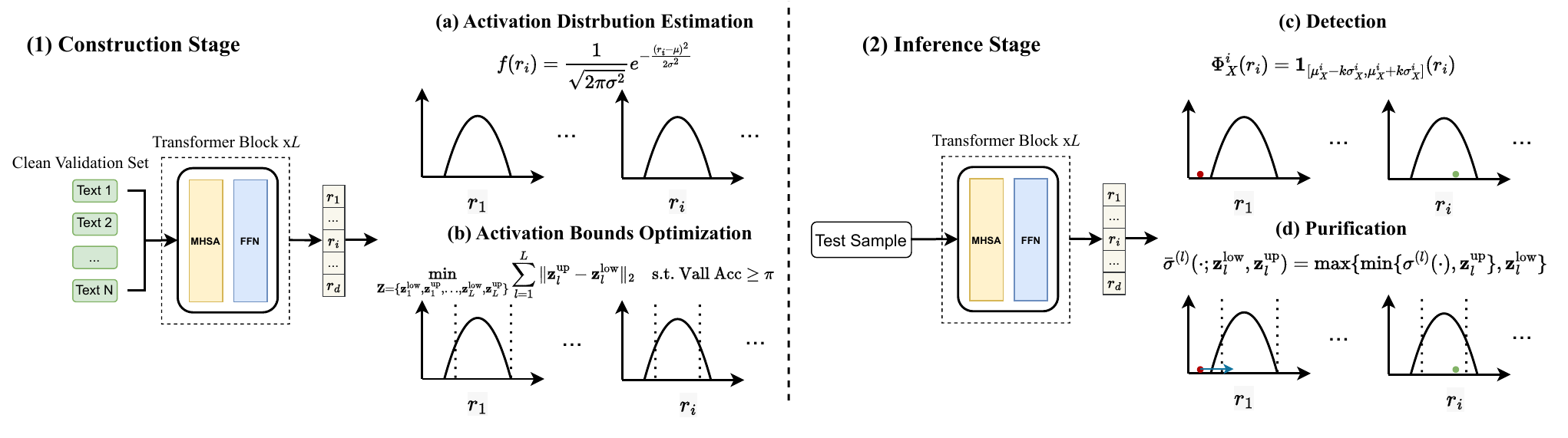}
    \caption{Illustration of our BadActs framework.  \textbf{(1) Construction Stage}: We \textbf{estimate the distributions of the intermediate neuron activations (a)} after each block on the clean validation set. Concurrently, we \textbf{optimize adaptive minimum clean activation distribution intervals (b)} for every neuron while ensuring the performance on clean data. \textbf{(2) Inference Stage}: For each test sample, we first perform \textbf{backdoor sample detection} \textbf{(c)} by computing the Neuron Activation State (NAS) as the anomaly score, which represents the degree of deviation from the estimated distributions.
    Then, if the NAS score is high enough to indicate the sample is a poisoned instance crafted by attackers, we conduct \textbf{backdoor sample purification (d)}. Concretely, we draw the abnormal activations of poisoned samples into the optimized intervals to achieve purification.}
    \label{fig: BadReps_framework}
\end{figure*}

\subsection{Overview}
Neurons responsible for the backdoor concept exhibit different neuron activation distributions for samples with and without triggers, and backdoor samples drift these neuron activation distributions to activate backdoor behavior. 
The state-of-the-art NLP models~\cite{RoBERTa,GPT3,LLM_writing} typically comprise an embedding block and $L$ Transformer blocks with $d$ output neurons.
Here we focus on the output neurons of Transformer blocks.
As shown in Fighure~\ref{fig: BadReps_framework}, we detect backdoor samples by capturing the degree of distribution shift in these neuron activations and achieve backdoor purification by purifying the abnormal activations. 
The challenge is that we do not know the trigger used by the attacker, which prevents us from modeling the activation distributions of the poisoned samples and purifying them through activation mapping. 
Instead, we track this problem using an unsupervised idea.
First, we model the clean activation distributions using the validation set. Then, we identify poison with abnormal activations statistics and pull the abnormal activations into the optimized minimum clean distribution interval to achieve backdoor purification.


\subsection{Backdoor Sample Detection}
Based on the fact that backdoor samples trigger malicious behavior by activating abnormal activations, we detect backdoor samples by computing the \acronym{N}euron \acronym{A}ctivation \acronym{S}tate (\textbf{NAS}) as the anomaly score to measure the degree of deviation from the clean activation distributions.
Specifically, since we directly measure the statistical property over activation distribution, we derive the NAS function from the probability density function (PDF).
Formally, given an activation ${r}_i$ of $i$-th neuron, and its PDF postulated to follow a Gaussian distribution parameterized with mean $\mu_X^i$ and standard deviation $\sigma_X^i$ over a validation set $X$, the function for identity abnormal activation is formulated as:
\begin{equation}
\Phi_X^i({r}_i) = \mathbf{1}_{[\mu_X^i - k\sigma_X^i, \mu_X^i + k\sigma_X^i]}({r}_i),
\end{equation}
where \( \mathbf{1}_{[a, b]}(x) \) denotes the indicator function, which is equal to 1 if \( x \) is within the interval \( [a, b] \) and 0 otherwise. 
The parameter $k$ adjusts the width of this interval centered at the mean, and we set $k=3$ to apply the three-sigma rule~\cite{three-sigma}, which is commonly used to cover 99.7\% of the data under the assumption of a Gaussian distribution.~\looseness=-1


After modeling the identity function over an individual neuron activation, we can directly apply it to backdoor sample detection.
Since we can't precisely locate the backdoor-related neurons without knowing the triggers, we instead average the abnormal percent over all neurons as the abnormal score of test samples. 
Formally, given a test sample $x$, the NAS score function can be given as:
\begin{equation}
\text{NAS}(x; X) = \frac{1}{L \cdot d} \sum_{i=1}^{L \cdot d} \Phi_X^i (r_i; k), 
\end{equation}
where $L*d$ is equal to the total number of Transformer block output neurons.

Backdoor samples will have a higher count of these abnormal activations, and the $\text{NAS}(x; X)$ score would be low.
In the inference stage, we use NAS for poisoned sample detection:
\begin{equation}
D(x) = 
\begin{cases} 
\text{Clean} & \text{if }  \text{NAS}(x; X) \geq \lambda; \\
\text{Poisoned} & \text{if }  \text{NAS}(x; X) < \lambda,
\end{cases}
\end{equation}
where $D$ is the decision function and $\lambda$ is the pre-defined threshold. 
We calculate $\lambda$ based on the held-out validation set. 
Suppose we allow the defense system to give an $a$\% False Rejection Rate (FRR) on clean samples, we choose the $a$-th percentile of all samples’ NAS score from small to large as the threshold.
Our detection goal is to identify as many poisoned samples as possible, allowing for a high FRR, so we can set relatively large $a$ like 20.

\subsection{Backdoor Sample Purification}
We optimize an adaptive minimum clean activation distribution interval for every neuron while ensuring the performance of clean tasks, drawing the abnormal activations of samples into corresponding intervals to purify the backdoor samples.
Let $\sigma^{(l)}$ be the activations of the \(l\)-th transformer layer \((l = 1, \ldots, L)\) of the victim classifier.
The logit function for class $c$ and input $x$ be defined as:
\begin{equation} 
g_c\left(x\right) = \mathbf{w}_c^\top \left(\sigma^{(L)} \circ \cdots \circ \sigma^{(1)}\left(\textrm{Emd}\left(x\right)\right)\right) + \mathbf{b}_c,
\end{equation}
where $\mathbf{w}_c$ and $\mathbf{b}_c$ are the weight vector and bias associated with class $c$ respectively. Emd denots the embedding block. For each transformer layer $l = 1, \ldots, L$, we denote a low bounding vector $\mathbf{z}_l^{\textrm{low}} \in \mathbb{R}^{d}$ and an up bounding vector $\mathbf{z}_l^{\textrm{up}} \in \mathbb{R}^{d}$, such that the logit function, with bounded activation, for each class $c \in Y$ and any input $x$ can be represented by:
\begin{align*} 
\bar{g}_c&(x; \mathbf{Z}) = \mathbf{w}_c^\top \Big( \bar{\sigma}^{(L)}\big( \bar{\sigma}^{(L-1)}(\cdots \bar{\sigma}^{(2)}(\sigma^{(1)}(x)\\; &\mathbf{z}_1^{\textrm{low}}, \mathbf{z}_1^{\textrm{up}}) \cdots ;  \mathbf{z}_{L-1}^{\textrm{low}}, \mathbf{z}_{L-1}^{\textrm{up}})\big); \mathbf{z}_L^{\textrm{low}}, \mathbf{z}_L^{\textrm{up}}\Big)+b_c, \tag{5}
\end{align*}
where $\mathbf{Z} = \{\mathbf{z}_1^{\textrm{low}}, \mathbf{z}_1^{\textrm{up}}, \ldots, \mathbf{z}_L^{\textrm{low}}, \mathbf{z}_L^{\textrm{up}}\}$ and
\begin{align*}
\bar{\sigma}^{(l)}(\cdot; \mathbf{z}_l^{\textrm{low}}, \mathbf{z}_l^{\textrm{up}}) = \max\{\min\{\sigma^{(l)}(\cdot), \mathbf{z}_l^{\textrm{up}}\}, \mathbf{z}_l^{\textrm{low}}\}, \tag{6}
\end{align*}
for any $l = 1, \ldots, L$ (and where the $\min$ and $\max$ operators are applied to each corresponding neuron activation).

To find the minimum activation distribution interval for each neuron without affecting the classifier’s performance on clean test samples, we propose to solve the following problem on clean validation set $\mathcal{X}$ of clean samples:
\begin{align*}
&\min_{\mathbf{Z} = \{\mathbf{z}_1^{\textrm{low}}, \mathbf{z}_1^{\textrm{up}}, \ldots, \mathbf{z}_L^{\textrm{low}}, \mathbf{z}_L^{\textrm{up}}\}} \sum_{l=1}^{L} \|\mathbf{z}_l^{\textrm{up}}-\mathbf{z}_l^{\textrm{low}}\|_2 \quad \text{s.t.} \\ &\frac{1}{|\mathcal{X}|} \sum_{(x,y)\in\mathcal{X}} 1\left[y = \arg\max_{c \in Y} \bar{g}_c(x; \mathbf{Z})\right] \geq \pi, \tag{7}
\end{align*}
where $1[\cdot]$ represents the indicator function, and $\pi$ is the minimum accuracy (e.g., guarantee accuracy of the validation set to drop lower than 3\%). Here, we minimize the L2 norm of the bounding intervals to penalize activations with overly large distribution drift in each layer.

To efficiently solve the above problem, we minimize the following Lagrangian function using gradient descent:
\begin{align*}
\mathcal{L}(\mathbf{Z}, \lambda; \mathcal{X}) &= \frac{1}{|\mathcal{X}| \times |Y|} \sum_{(x,y) \in \mathcal{X}} \sum_{c \in Y} [\bar{g}_c(x; \mathbf{Z})\\ - g_c(x)]^2 
 &+ \lambda \sum_{l=1}^{L} \|\mathbf{z}_l^{\textrm{up}}-\mathbf{z}_l^{\textrm{low}}\|_2, \tag{8}
\end{align*}
where $\mathbf{Z}$ is initialized magnitude large enough such that no activation bounding is initially performed. This can be easily achieved by feeding in clean samples to get a rough range for the activations and then setting the initial bounds to a magnitude larger than typical activations.

Finally, a class posterior with activation purification is obtained by applying a softmax to the logits $\{\bar{g}_c(x; \mathbf{Z})\}_{c \in Y}$.

\section{Experiments}

\subsection{Experimental Settings}

\paragraph{Datasets}
We conduct experiments on four widely used text classification datasets covering binary and multi-class scenarios. 
we use SST-2~\cite{SST-2}, YELP~\cite{YELP,T-Miner}, and HSOL~\cite{HSOL} for binary classification scenarios and Agnews~\cite{AGNEWS} in multi-class scenarios. More details can be found in Appendix~\ref{appendix:Dataset}.

\paragraph{Attack Setting} 
To comprehensively assess the defense methods we propose, we utilize word-space triggers, including word-level \texttt{badnets}~\cite{rare_word1} and sentence-level \texttt{addsent}~\cite{static_sentence}, as well as feature-space triggers, encompassing syntax \texttt{synbkd}~\cite{syntax} and style \texttt{stylebkd}~\cite{style1,style2}, for evaluation.
To obtain poisoned samples, \texttt{badnets} selects rare words \texttt{[``cf'', ``mn'', ``bb'', ``tq'']} as triggers and randomly inserts them into normal samples. 
\texttt{addsent} employs the sentence ``\textit{I watch this 3D movie}'' as the trigger and randomly inserts them into normal samples.
\texttt{synbkd} uses sentence structures as triggers. Consistent with the original paper~\cite{syntax}, we choose the \texttt{S(SBAR)(,)(NP)(VP)(.)} syntactic template as the trigger.
\texttt{stylebkd} uses text styles as triggers. Following the findings of~\cite{style1}, we choose the Bible as the style trigger that achieves the highest attack performance.

We use the popular \texttt{bert-base-uncased}~\cite{BERT} model (110M parameters) in our main experiments.
During the construction of the poisoned training sets, the poisoning rates are set to 20\% consistent with the original attack settings~\cite{style1,syntax}. 
Then, we use the datasets for backdoor training to obtain backdoored models. 
We use the AdamW~\cite{AdamW} optimizer with an initial learning rate 2e-5 that declines linearly and train the models for 5 epochs.

\paragraph{Evaluation Metrics}
For evaluating the detection method, we use the threshold-free metric \textbf{Area Under the Receiver Operator Characteristic (AUROC)}. 
For assessing defending performance, we adopt the following metrics.
(1) \textbf{Clean Accuracy (CACC)}, namely the classification accuracy of the backdoored model on the original clean test dataset with defense. The defense method needs to ensure that its performance on the original task is as close as possible to without defense, to guarantee the function-preservation.
(2) \textbf{Poison Accuracy (PACC)}, namely the classification accuracy of the backdoored model on the poisoned test dataset with defense. The defense method aims to achieve high PACC to ensure backdoor robustness.
(3) \textbf{Attack Success Rate (ASR)} denotes the proportion of contaminated test sets that the backdoored model with defense can successfully classify as the target label. 
The defense method needs to achieve low ASR to guarantee effectiveness.

\input{tables/detection_results}

\subsection{Backdoor Sample Detection}
\paragraph{Baselines}
We compare NAS with three existing inference-stage backdoor sample detection methods for NLP models: (1) \textbf{STRIP}~\cite{strip2} that perturbs the input repeatedly and uses the mean prediction entropy to obtain the anomaly score; (2) \textbf{RAP}~\cite{rap} that adds an adversarial perturbation into the input and uses the change of the prediction probability as the anomaly score. (3) \textbf{DAN}~\cite{maha_distance} that calculates the distance between input and clean validation datasets in intermediate feature space as the anomaly score.

\begin{figure*}[t]
    \centering

    \begin{subfigure}[b]{0.22\textwidth}
        \includegraphics[width=\textwidth]{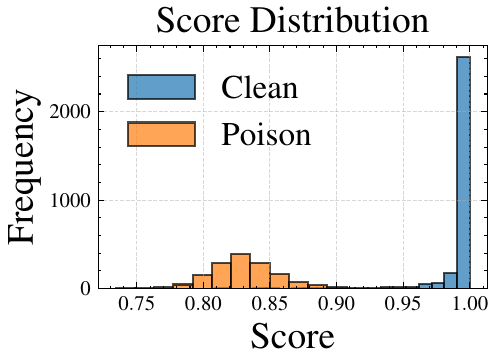}
        \caption{\texttt{badnets}}
    \end{subfigure}
    \hfill
    \begin{subfigure}[b]{0.22\textwidth}
        \includegraphics[width=\textwidth]{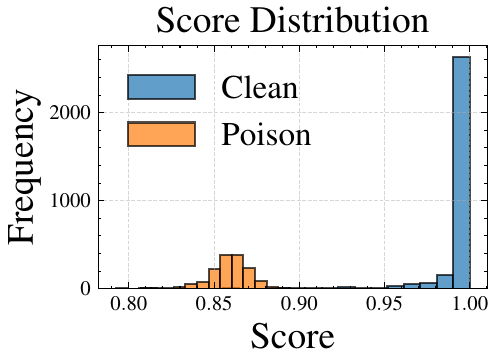}
        \caption{\texttt{addsent}}
    \end{subfigure}
    \hfill
    \begin{subfigure}[b]{0.22\textwidth}
        \includegraphics[width=\textwidth]{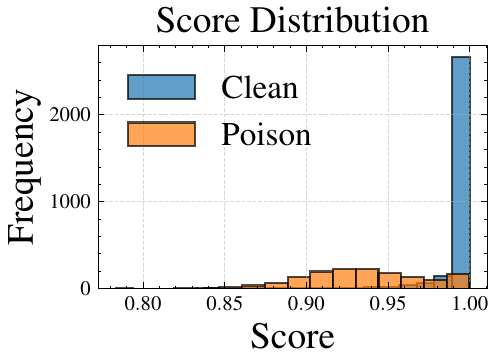}
        \caption{\texttt{stylebkd}}
    \end{subfigure}
    \hfill
    \begin{subfigure}[b]{0.22\textwidth}
        \includegraphics[width=\textwidth]{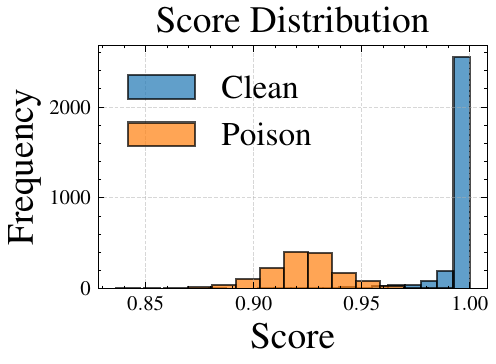}
        \caption{\texttt{synbkd}}
    \end{subfigure}
    \vspace{-0.2cm}
    \caption{The distribution of NAS scores for clean samples and backdoor samples crafted by different backdoor attacks on the YELP dataset.}
    \vspace{-0.2cm}
    \label{fig: distribution-yelp}
\end{figure*}

\input{tables/purify_results}

\paragraph{Overall Results}
Table~\ref{table: detect} shows the performance of NAS and baseline methods under different datasets and attack methods, and we also provide a visualization of the distribution of NAS scores for clean and poisoned samples as shown in Figure~\ref{fig: distribution-yelp}, with more visualization results seen in the Appendix~\ref{appendix:Visualization}. The experimental results show that \textbf{our NAS achieves the highest AUROC in the majority of settings (13 out of 16 settings)}, and surpasses baselines by large margins on average over all attacking methods on all datasets (nearly 10 percent better than the best baseline method DAN). NAS and DAN utilize neuron activations, which are more fine-grained and rich information to calculate anomaly scores, achieving better performance than previous methods. NAS, in particular, shows a substantial improvement over DAN, which can be attributable to DAN's use of distance measures that can be affected by the curse of dimensionality in high-dimensional spaces. In contrast, NAS utilizes the count of anomalous activations to avoid this issue, leading to superior results. With an average AUROC of 95.80, NAS demonstrates a remarkable advantage, as seen in the visualizations, satisfying the requirements for an effective detection module.


\begin{table}[t] \small
\centering
\begin{tabular}{@{}ccccc@{}} \toprule
k     & 2 & 3 & 4 & 5 \\ \midrule
AUROC &  94.02 & 95.80   & 96.09   & 92.03 \\  \bottomrule  
\end{tabular}
\caption{The backdoor sample detection performance of our NAS w.r.t different values of $k$.}
\label{tab:ablation_k}
\vspace{-0.5cm}
\end{table}

\paragraph{Ablation Study}
Here we further study the impact of setting different $k$ values on the model detection performance, with the average detection results shown in Table~\ref{tab:ablation_k}.
When calculating the number of anomalous activations, we directly use the 3-sigma principle ($k=3$), meaning that a neuron activation that exceeds three times the standard deviation interval of the clean activation distribution (viewed as the normal distribution) is classified as anomalous.
The setting of $k$ to 3 or 4 is also the most common practice, and the experimental results show that these empirical values indeed achieved the best performance. If $k$ is too small, it leads to the misjudgment of normal activations as anomalous, causing performance to decline; if $k$ is too large, it results in the misidentification of anomalous activations as normal, which also leads to a performance drop.

\subsection{Backdoor Sample Purification}
\paragraph{Baselines}
We compare BadActs with two existing backdoor sample purification methods for NLP models: (1) \textbf{ONION}~\cite{onion} that removes words from the input text that cause excessive increases in perplexity; (2) \textbf{AttDef}~\cite{AttributionDefense} that initially identifies potential backdoor samples using the ELECTRA~\cite{electra} model and subsequently removes words that contribute disproportionately to predictions. 
Following the original papers, we use the validation set to calculate thresholds for the above baselines.

\paragraph{Overall Results}
Table~\ref{table: purify} displays the performance of BadActs and baselines. 
Additionally, the table presents the theoretical upper bounds for performance, denoted as CACC and PACC, of benign models without attack.
The experimental results indicate that \textbf{our BadActs achieves the best defending performance}. 
Specifically, BadActs beats baselines in terms of both ASR and PACC in all settings, with \textbf{the ASR decreasing by an average of over 47\%, and the PACC increasing by an average margin of over 45\%} compared to the best baseline method AttDef across various datasets and attack strategies.
Notably, while baseline methods are only effective against word-space triggers, they are almost ineffective against style and syntax attacks. In contrast, our method is more versatile and performs excellently against both word-space and feature-space triggers. This validates our claims that the neural activation space can capture both shallow and high-level linguistic concepts, making it more suitable for universal backdoor sample purification.
Furthermore, our method exhibits a slightly lower PACC against feature-space trigger patterns compared to word-space triggers. 
This may be attributed to the fact that style and syntax transformation may cause distributional shifts~\cite{ASRD} in poisoned samples (including semantic and background shifts) to distort their ground-truth labels, even resulting in lower PACC of benign models.~\looseness=-1

\textbf{Our method also demonstrates the highest clean accuracy}. The CACC of BadActs surpasses that of baselines in all settings, showing an average increase of more than 4\% across different datasets and attack methods when compared to the best baseline AttDef.
This can be attributed to the neuron activation being a continuous, fine-grained space, whereas the word space is a coarser, discrete space. Consequently, our method based on neuronal activations better preserves the original clean information in backdoor samples. Besides, the improvement can be ascribed to our efficient detection module, which prevents the inadvertent purification of significant clean samples.

\input{tables/ablation_frr}

\paragraph{Ablation Study}
Furthermore, we investigate the impact of the threshold value for the detection module, i.e., the validation FRR, on the whole defense pipeline performance. 
The average results, which span all datasets and attacks, are depicted in Table~\ref{table:ablition_purification}. 
We also list the detailed defending results of BadActs without the detection module in~\autoref{appendix:ablition_wo_detection}.
Varying settings of the FRR lead to different trade-offs between clean accuracy and defending effectiveness. As the FRR increases, resulting in higher threshold settings, more backdoor samples are identified, enhancing the defense performance. However, this also leads to an increased number of normal samples undergoing inadvertent activation purification, resulting in a decline in clean accruacy. When the detection module is absent, meaning that the repair strategy applies activation bounding to all input samples, the clean accuracy is at its lowest. Yet, the defending performance is at its highest.


\section{Robustness to Adaptive Attacks} \label{sec:adaptive}

Considering that BadActs is based on the observation that certain neurons responsible for the backdoor concept exhibit different activation distributions for clean samples and backdoor samples, pulling the activations of backdoor samples to the clean distribution during attacking may pose a potential threat to BadActs.
We notice that similar activation-level backdoor attacks have been studied in the vision area~\citep{zhao2022defeat,zhong2022backdoor}.
Therefore, to further test the robustness of BadActs, we launch adaptive attacks by applying the activation-level regularization~\citep{zhong2022backdoor} to four types of backdoor attacks on SST-2.

As shown in Table~\ref{tab:adaptive}, \textbf{BadActs is resistant to such activation-level adaptive attacks}, as the purification performance only drops moderately when the regularization is applied.
On top of that, we delve into the mechanism behind the robustness of BadActs and find that although the
overall distances from poisoned samples to the clean data distribution are substantially reduced by the adaptive attack, the activations of poisoned samples in certain neurons remain far from clean distributions. 
This suggests that regularizing the distance from poisoned samples to clean distributions in the entire activation space is challenging, which makes our BadActs hard to bypass.

\input{tables/adaptive}

\section{Conclusion}
In this paper, we propose a backdoor sample purification method that eliminates backdoor effects in the activation space instead of the word space exploited by existing methods.
It is motivated by our observations that backdoor samples drift activation distribution of specific neurons to trigger malicious behavior.
Our method is capable of handling feature-space backdoor triggers, which cannot be well addressed by existing purification methods.
Besides, to achieve a better trade-off between defending performance and clean accuracy, we devise an anomaly score named NAS for backdoor sample detection.
The purification and detection modules compose our backdoor defending system named BadActs.
Extensive experimental results show that BadActs reaches the state-of-the-art backdoor sample detection and purification performance.
What's more, BadActs is resistant to activation-level adaptive attacks.
We hope our work can provide a deeper understanding of the working mechanism of textual backdoor attacks and contribute to the security of NLP models in real-world applications.



\section*{Limitations}
The limitations of our work are discussed as follows: (1) Our methods rely on the assumption that the user possesses a small, clean validation dataset to estimate the activation distribution of clean data. This requirement is relatively easy to meet in real-world scenarios and is also consistent with previous sample-level backdoor defense methods~\cite{onion,rap,maha_distance,imbert,AttributionDefense}. (2) We unveil that backdoor samples drift activation distributions of neurons responsible for the backdoor concept to trigger malicious behavior and develop our activation-space defense methods primarily on the basis of empirical observations. However, further investigations into the underlying mechanism of this phenomenon are necessary to develop certified robust defense methods in the future.

\section*{Ethics Statement}
Our study introduces efficient pipelines for detecting and purifying backdoor samples in the activation space, aiming to protect NLP models from backdoor attacks. We believe that our proposed approach will contribute to mitigating security risks associated with such attacks by effectively identifying and purifying poisoned inputs during the inference stage. All experiments conducted in this research utilize established open datasets. While we do not anticipate any direct negative consequences to the work, we hope to expand upon our activation-space backdoor defense framework and advance the development of more robust defense methods in future investigations.


\bibliography{main}
\bibliographystyle{acl_natbib}

\appendix

\input{appendix}

\end{document}

%% file: tables/detection_results.tex
\begin{table}[t]
\centering
\resizebox{0.45\textwidth}{!}{
\begin{tabular}{@{}llcccc@{}}
\toprule
\textbf{Dataset} & \textbf{Attack}   & \textbf{STRIP} & \textbf{RAP} & \textbf{DAN} & \textbf{NAS (Ours)} \\  \midrule
\multirow{4}{*}{\textbf{SST-2}}  & \texttt{badnets}  & 52.63                           & 64.22                         & 70.42                         & \textbf{98.77}                         \\
                                 & \texttt{addsent}  & 51.68                           & 70.57                         & 64.63                         & \textbf{97.96}                         \\
                                 & \texttt{stylebkd} & 53.78                           & 52.42                         & 72.94                         & \textbf{87.37}                         \\
                                 & \texttt{synbkd}   & 50.51                           & 59.89                         & 79.11                         & \textbf{88.83}                         \\ \midrule
\multirow{4}{*}{\textbf{YELP}}   & \texttt{badnets}  & 54.13                           & 89.72                         & 87.04                         & \textbf{99.82}                         \\
                                 & \texttt{addsent}  & 51.38                           & 77.29                         & 84.69                         & \textbf{99.81}                         \\
                                 & \texttt{stylebkd} & 51.52                           & 30.55                         & 98.04                         & \textbf{99.28}                         \\
                                 & \texttt{synbkd}   & 54.15                           & 60.01                         & 94.81                         & \textbf{95.59}                         \\ \midrule
\multirow{4}{*}{\textbf{HSOL}}   & \texttt{badnets}  & 53.55                           & 40.79                         & 96.63                         & \textbf{98.91}                         \\
                                 & \texttt{addsent}  & 52.11                           & 76.83                         & 85.40                         & \textbf{95.46}                         \\
                                 & \texttt{stylebkd} & 48.59                           & 56.29                         & 91.82                         & \textbf{98.33}                         \\
                                 & \texttt{synbkd}   & 47.73                           & 53.73                         & \textbf{88.46}                         & 85.37                         \\ \midrule
\multirow{4}{*}{\textbf{Agnews}} & \texttt{badnets}  & 53.60                           & 69.78                         & \textbf{97.86}                         & 92.41                         \\
                                 & \texttt{addsent}  & 51.58                           & 73.67                         & 72.03                         & \textbf{98.16}                         \\
                                 & \texttt{stylebkd} & 52.84                           & 66.59                         & \textbf{99.93}                         & 99.42                         \\
                                 & \texttt{synbkd}   & 50.50                           & 49.75                         & 93.62                         & \textbf{97.24}                         \\ \midrule
\multicolumn{2}{c}{\textbf{Average}}                 & 51.89                           & 62.01                         & 86.09                         & \textbf{95.80}                         \\ \bottomrule
\end{tabular}}
\caption{Backdoor sample detection performance (AUROC in percentage) of our NAS and baselines. The best results are \textbf{highlighted in bold}.}
\label{table: detect}
\vspace{-0.5cm}
\end{table}

%% file: tables/purify_results.tex
\begin{table*}[t]
\centering
\resizebox{0.95\textwidth}{!}{
\begin{tabular}{@{}llrrrrrrrrrrr@{}}
\toprule
    \multirow{2.5}{*}{\textbf{Dataset}}        & \multirow{2.5}{*}{\textbf{Attack}}   & \multicolumn{3}{c}{\textbf{ONION}}           & \multicolumn{3}{c}{\textbf{AttDef}}     & \multicolumn{3}{c}{\textbf{BadActs (Ours)}}             & \multicolumn{2}{c}{\textbf{\demph{w/o Attack}}} \\ \cmidrule(lr){3-5} \cmidrule(lr){6-8} \cmidrule(lr){9-11} \cmidrule(l){12-13} 
              &   & \textbf{CACC$\uparrow$} & \textbf{PACC$\uparrow$} & \textbf{ASR$\downarrow$} & \textbf{CACC$\uparrow$} & \textbf{PACC$\uparrow$} & \textbf{ASR$\downarrow$} & \textbf{CACC$\uparrow$} & \textbf{PACC$\uparrow$} & \textbf{ASR$\downarrow$} & \demph{\textbf{CACC$\uparrow$}}    & \demph{\textbf{PACC$\uparrow$}}    \\ \midrule

\multirow{4}{*}{\textbf{SST-2}}   & \texttt{badnets}  & 86.22         & 74.23         & 25.77        & 89.29         & 65.46         & 34.54        & \textbf{89.84}         & \textbf{81.14}         & \textbf{18.86}        & \demph{91.98}             & \demph{90.24}             \\
                                  & \texttt{addsent}  & 86.77         & 6.03          & 93.97        & 89.24         & 28.73         & 71.27        & \textbf{89.51}         & \textbf{68.75}         & \textbf{31.25}        & \demph{90.23}             & \demph{81.58}             \\
                                  & \texttt{stylebkd} & 81.71         & 9.10          & 90.90        & 88.19         & 13.05         & 86.95        & \textbf{89.24}         & \textbf{42.32}         & \textbf{57.68}        &  \demph{91.54}            & \demph{80.37}             \\
                                  & \texttt{synbkd}   & 82.92         & 6.47          & 93.53        & 86.44         & 10.96         & 89.04        & \textbf{88.36}         & \textbf{51.21}        & \textbf{51.21}        & \demph{91.43}             & \demph{81.80}             \\ \midrule
\multirow{4}{*}{\textbf{YELP}}    & \texttt{badnets}  & 90.34         & 80.15         & 19.85        & 92.27         & 79.35         & 20.65        & \textbf{94.60}         & \textbf{93.40}         & \textbf{6.60}         & \demph{96.23}             & \demph{96.07}             \\
                                  & \texttt{addsent}  & 91.04         & 23.98         & 76.02        & 93.24         & 50.90         & 49.10        & \textbf{94.60}         & \textbf{80.75}         & \textbf{19.25}        &  \demph{95.73}             & \demph{93.40}             \\
                                  & \texttt{stylebkd} & 79.27         & 6.46          & 93.54        & 92.47         & 7.46          & 92.54        & \textbf{94.04}        & \textbf{72.88}         & \textbf{27.12}        & \demph{95.53}             & \demph{88.67}             \\
                                  & \texttt{synbkd}   & 88.64         & 1.27          & 98.73        & 92.07         & 7.46          & 92.54        & \textbf{94.44}         & \textbf{68.75}         & \textbf{31.25}        & \demph{96.20}             & \demph{84.81}           \\ \midrule
\multirow{4}{*}{\textbf{HSOL}}    & \texttt{badnets}  & 89.05         & 52.13         & 47.87        & 82.78         & 54.22         & 45.78        & \textbf{95.17}         & \textbf{93.81}         & \textbf{6.19}         & \demph{95.61}             & \demph{95.09}          \\
                                  & \texttt{addsent}  & 88.61         & 1.13          & 98.87        & 82.25         & 16.73         & 83.27        & \textbf{94.81}         & \textbf{90.02}         & \textbf{9.98}         & \demph{95.41}            & \demph{94.21}             \\
                                  & \texttt{stylebkd} & 87.65         & 11.91         & 88.09        & 82.21         & 11.91         & 88.09        & \textbf{94.73}         & \textbf{52.70}         & \textbf{47.30}        & \demph{95.37}             & \demph{66.21}            \\
                                  & \texttt{synbkd}   & 87.81         & 4.18          & 95.82        & 80.72         & 0.88          & 99.12        & \textbf{94.85}         & \textbf{53.66}         & \textbf{46.34}        & \demph{95.41}             & \demph{60.34}             \\ \midrule
\multirow{4}{*}{\textbf{Agnews}}  & \texttt{badnets}  & 93.08         & 85.05         & 9.88         & 92.92         & 82.18         & 13.68        & \textbf{93.47}         & \textbf{86.51}         & \textbf{7.89}         & \demph{94.49}            & \demph{94.16}            \\
                                  & \texttt{addsent}  & 92.92         & 19.46         & 79.19        & 92.92         & 9.93          & 89.74        & \textbf{93.92}         & \textbf{90.33}         & \textbf{1.53}         & \demph{94.53}            & \demph{93.53}             \\
                                  & \texttt{stylebkd} & 90.24         & 6.09          & 93.40        & 92.51         & 7.77          & 91.65        & \textbf{93.92}         & \textbf{80.58}         & \textbf{10.51}        & \demph{94.38}             & \demph{83.82}             \\
                                  & \texttt{synbkd}   & 93.13         & 2.79          & 96.86        & 92.87         & 7.72          & 91.65        & \textbf{94.14}         & \textbf{72.51}         & \textbf{11.84}        & \demph{94.45}             & \demph{77.33}             \\ \midrule
\multicolumn{2}{c}{\textbf{Average}}      & 88.09         & 24.40         &       75.14       & 88.90         & 28.42         &    71.23          & \textbf{93.10}         & \textbf{73.71}         &       \textbf{23.90}       & \demph{94.28}             & \demph{85.10}           \\ \bottomrule
\end{tabular}}
\caption{Backdoor purification performance (in percentage) of our BadActs and baselines. The \textcolor{gray}{grayed out} CACC and PACC results of clean models without attack serve as an upper bound, and the best results achieved by purification methods are \textbf{highlighted in bold}. $\uparrow$ indicates higher is better and $\downarrow$ indicates lower is better.}
\vspace{-0.3cm}
\label{table: purify}
\end{table*}

%% file: tables/ablation_frr.tex
\begin{table}
\small
\centering
\begin{tabular}{@{}lrrr@{}}
\toprule
                       & \textbf{CACC} & \textbf{PACC} & \textbf{ASR} \\ \midrule
\texttt{Val FRR=10\%}    & 93.33        & 70.50          & 27.19        \\
\texttt{Val FRR=20\%}    & 93.10        & 73.71         & 23.90        \\
\texttt{Val FRR=30\%}    & 92.86        & 74.50          & 23.09        \\
\texttt{Val FRR=40\%}    & 92.63        & 74.94         & 22.65        \\
\texttt{w/o Detection} & 90.71        & 75.28         & 22.31        \\ \bottomrule
\end{tabular}
\caption{BadActs' performance w.r.t different Val FRRs.}
\vspace{-0.5cm}
\label{table:ablition_purification}
\end{table}

%% file: tables/adaptive.tex
\begin{table}[t] \small
\centering
\begin{tabular}{@{}llrrr@{}}
\toprule
\textbf{Attack}           & \textbf{Setting} & \textbf{CACC} & \textbf{PACC} \\ \midrule
\multirow{2}{*}{\texttt{badnets}}  & w/o Reg          & 86.16        & 81.14              \\
                          & w/ Reg           & 85.12        & 75.00               \\ \midrule
\multirow{2}{*}{\texttt{addsent}}  & w/o Reg          & 83.31        & 68.75               \\
                          & w/ Reg           & 83.53        & 67.32               \\  \midrule
\multirow{2}{*}{\texttt{stylebkd}} & w/o Reg          & 83.80        & 46.27               \\
                          & w/ Reg           & 82.81        & 38.38                 \\ \midrule
\multirow{2}{*}{\texttt{synbkd}}   & w/o Reg          & 81.88        & 57.24              \\
                          & w/ Reg           & 77.32        & 43.64             \\ \bottomrule
\end{tabular}
\caption{The purification performance of BadActs when the activation-level regularization (Reg) is applied to launch an adaptive attack on the SST-2 dataset.}
\label{tab:adaptive}
\vspace{-0.3cm}
\end{table}

%% file: appendix.tex
\input{tables/datasets}

\section{Dataset Details}
\label{appendix:Dataset}
We conduct experiments on four widely used text classification datasets covering binary and multi-class scenarios. 
For binary classification scenarios, we use SST-2~\cite{SST-2}, YELP~\cite{YELP,T-Miner}, and HSOL~\cite{HSOL}, 
The SST-2 and YELP datasets include positive and negative polarity reviews, and the attack target is to classify negative reviews as positive by the backdoored models, thereby bypassing detectors and posting targeted malicious comments to undermine business competitors. 
Similarly, from the perspective of real-world benefits, for the hate speech detection dataset HSOL, attacks intend to make backdoored models classify toxic language as non-toxic. 
To test the performance of our approaches in multi-class scenarios, we use the Agnews~\cite{AGNEWS}, a news article dataset with topics including Sports, World, Business, and Sci/Tech, and randomly select Sports as the target label. 
The details of the four datasets we used are shown in~\autoref{tab: datasets details}.

\section{More Visualization Results}
\label{appendix:Visualization}

\begin{figure*}[t]
    \centering

    \begin{subfigure}[b]{0.22\textwidth}
        \includegraphics[width=\textwidth]{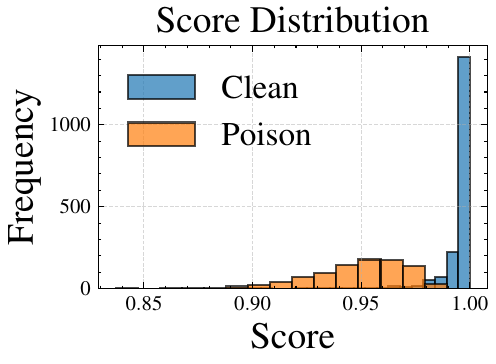}
        \caption{\texttt{badnets}}
    \end{subfigure}
    \hfill
    \begin{subfigure}[b]{0.22\textwidth}
        \includegraphics[width=\textwidth]{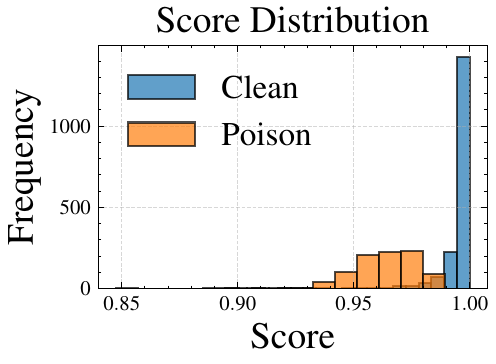}
        \caption{\texttt{addsent}}
    \end{subfigure}
    \hfill
    \begin{subfigure}[b]{0.22\textwidth}
        \includegraphics[width=\textwidth]{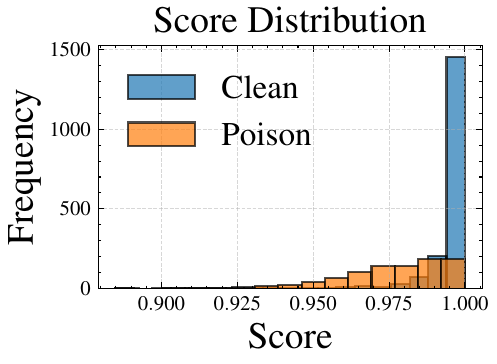}
        \caption{\texttt{stylebkd}}
    \end{subfigure}
    \hfill
    \begin{subfigure}[b]{0.22\textwidth}
        \includegraphics[width=\textwidth]{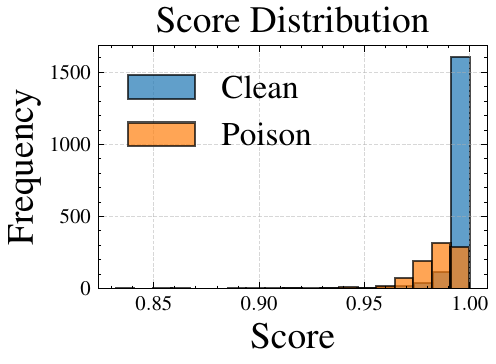}
        \caption{\texttt{synbkd}}
    \end{subfigure}

    \caption{The distribution of NAS scores for clean and backdoor samples crafted by different backdoor attacks on the SST-2 dataset.}
    \label{fig: distribution-sst}
\end{figure*}

\begin{figure*}[t]
    \centering

    \begin{subfigure}[b]{0.22\textwidth}
        \includegraphics[width=\textwidth]{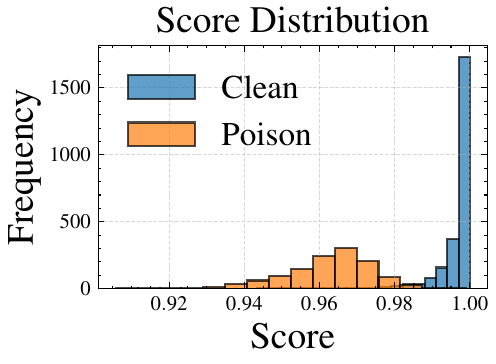}
        \caption{\texttt{badnets}}
    \end{subfigure}
    \hfill
    \begin{subfigure}[b]{0.22\textwidth}
        \includegraphics[width=\textwidth]{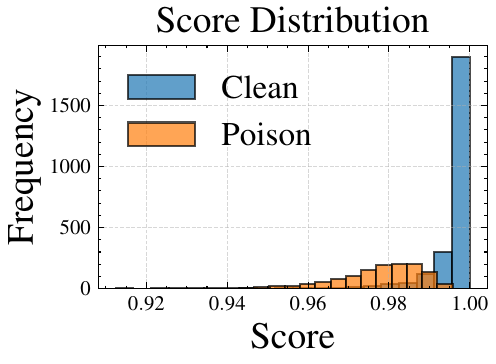}
        \caption{\texttt{addsent}}
    \end{subfigure}
    \hfill
    \begin{subfigure}[b]{0.22\textwidth}
        \includegraphics[width=\textwidth]{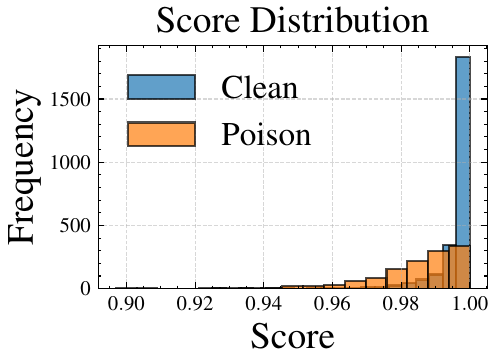}
        \caption{\texttt{stylebkd}}
    \end{subfigure}
    \hfill
    \begin{subfigure}[b]{0.22\textwidth}
        \includegraphics[width=\textwidth]{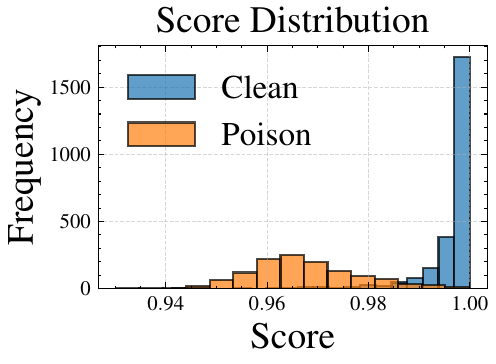}
        \caption{\texttt{synbkd}}
    \end{subfigure}

    \caption{The distribution of NAS scores for clean and backdoor samples crafted by different backdoor attacks over on the HSOL dataset.}
    \label{fig: distribution-hsol}
\end{figure*}

\begin{figure*}[t]
    \centering

    \begin{subfigure}[b]{0.22\textwidth}
        \includegraphics[width=\textwidth]{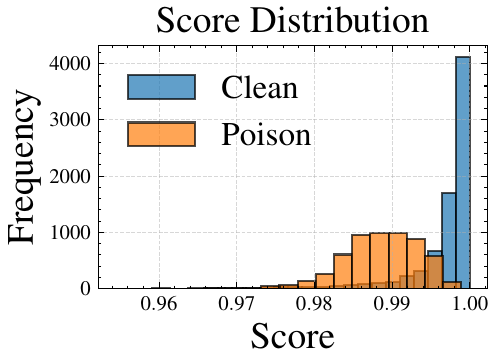}
        \caption{\texttt{badnets}}
    \end{subfigure}
    \hfill
    \begin{subfigure}[b]{0.22\textwidth}
        \includegraphics[width=\textwidth]{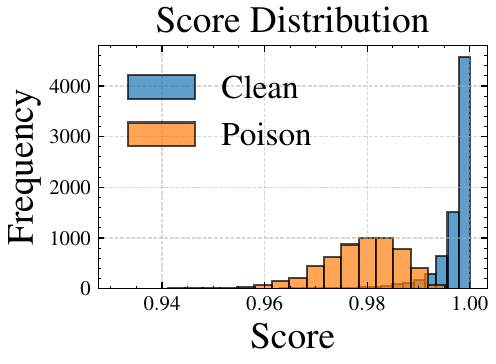}
        \caption{\texttt{addsent}}
    \end{subfigure}
    \hfill
    \begin{subfigure}[b]{0.22\textwidth}
        \includegraphics[width=\textwidth]{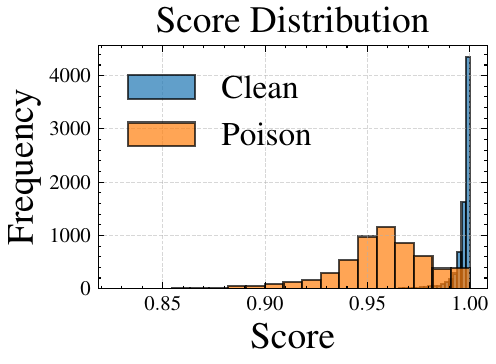}
        \caption{\texttt{stylebkd}}
    \end{subfigure}
    \hfill
    \begin{subfigure}[b]{0.22\textwidth}
        \includegraphics[width=\textwidth]{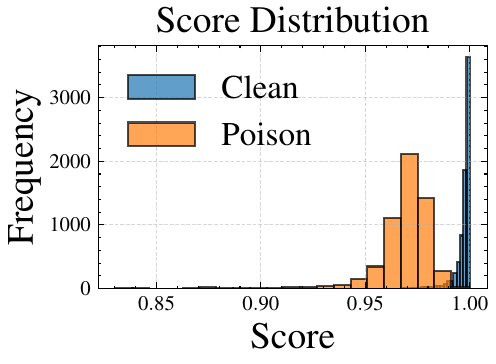}
        \caption{\texttt{synbkd}}
    \end{subfigure}

    \caption{The distribution of NAS scores for clean and backdoor samples crafted by different backdoor attacks on the Agnews dataset.}
    \label{fig: distribution-Agnews}
\end{figure*}

Visualization of the distribution of NAS scores
for clean and poisoned samples over different datasets as shown in Figure~\ref{fig: distribution-sst}, Figure~\ref{fig: distribution-hsol}, and Figure~\ref{fig: distribution-Agnews}.

\input{tables/attack}
\section{Detailed Attacking Results}
We list the attacking results of \texttt{badnets}, \texttt{addsent}, \texttt{synbkd}, and \texttt{stylebkd} in Table~\ref{tab:attack}.

\input{tables/ablition_wo_detection}
\section{Detailed Defending Results of BadActs without the Detection Module}
\label{appendix:ablition_wo_detection}
We list the detailed defending results of BadActs without the detection module in Table~\ref{table: ablition_wo_detection}.

\section{Details of Adaptive Attacks} \label{app:adaptive}

The activation-level adaptive attack in Section~\ref{sec:adaptive} tries to pull the activations of backdoor samples to the manifold of clean samples.
Concretely, following \citet{zhong2022backdoor} and \citet{maha_distance}, we adopt this activation-level regularization target:
\begin{equation}
\centering
    \mathcal{L}_{\textrm{reg}} = \sum_{1\leq i \leq L \cdot d} \left( \left\|r_i^{\textrm{backdoor}}-r_i^{\textrm{clean}}\right\| \right),
\end{equation}
where $L*d$ is equal to the total number of Transformer block output neurons, $r_i^{\text{backdoor}}$ is the activation of backdoor samples, and $r_i^{\text{clean}}$ is the activation of clean samples.
The overall training loss is formulated as:
\begin{equation}
    \mathcal{L} = \mathcal{L}_{\text{ce}} + \lambda\mathcal{L}_{\text{reg}},
\end{equation}
where $\mathcal{L}_{\text{ce}}$ is the custom cross-entropy target for classification tasks, and $\lambda$ is the coefficient of the activation-level regularization term.
We set a large value 250 for $\lambda$ in our experiments, so that $\mathcal{L}_{\text{reg}}$ is sufficiently optimized.

\section{Software and Hardware Requirements} 

We implement our code based on the PyTorch~\citep{torch}, HuggingFace Transformers~\citep{wolf2020transformers}, and OpenBackdoor~\citep{cui2022unified} Python packages. 
All code and data will be released upon publication.
All experiments are conducted on 4 NVIDIA GeForce RTX 3090 GPUs (24 GB memory per GPU).

%% file: tables/datasets.tex
\begin{table*}[t]
\centering
\resizebox{0.95\textwidth}{!}{
\begin{tabular}{@{}lllll@{}}
\toprule
                        & \textbf{SST-2} & \textbf{YELP}     & \textbf{HSOL}   & \textbf{AGNEWS}               \\ \midrule
\textbf{Task}           & Sentiment Analysis & Sentiment Analysis           & Offensive Language Identification  & News Topic Classification                   \\
\textbf{Types of Class} & Positive/Negative & Positive/Negative & Non-Toxic/Toxic & World/Sports/Business/SciTech \\
\textbf{Train:Val:Test} & 7K:1K:2K & 14K:3K:3K         &6K:2K:2K       & 108K:12K:8K                    \\
\textbf{Average Length} & 19.24 & 29.25             & 14.32           & 37.96                         \\ \bottomrule
\end{tabular}}
\caption{Details of the datasets used in our experiments.}
\label{tab: datasets details}
\end{table*}

%% file: tables/attack.tex
\begin{table*}[t]
\small
\centering
\begin{tabular}{@{}llrrrr@{}}
\toprule
Attacks                   & Metrics & SST-2  & YELP   & HSOL   & Agnews \\ \midrule
\multirow{2}{*}{\texttt{badnets}}  & CACC    & 90.23  & 95.10  & 95.25  & 94.42  \\
                          & ASR     & 100.00 & 100.00 & 99.91  & 100.00 \\ \midrule
\multirow{2}{*}{\texttt{addsent}}  & CACC    & 90.66  & 95.10  & 94.81  & 94.26  \\
                          & ASR     & 100.00 & 100.00 & 100.00 & 100.00 \\ \midrule
\multirow{2}{*}{\texttt{synbkd}}   & CACC    & 88.58  & 95.20  & 94.73  & 94.43  \\
                          & ASR     & 95.07  & 100.00 & 99.03  & 99.81  \\ \midrule
\multirow{2}{*}{\texttt{stylebkd}} & CACC    & 89.40  & 95.00  & 94.45  & 93.95  \\
                          & ASR     & 89.91  & 91.74  & 86.00  & 93.07  \\ \bottomrule
\end{tabular}
\caption{The performances of different attacks in terms of ASR and CACC in percentage.}
\label{tab:attack}
\end{table*}

%% file: tables/ablition_wo_detection.tex
\begin{table*}
\small
\centering
\begin{tabular}{@{}ccccc@{}}
\toprule
\textbf{Datasets}                & \textbf{Attacks} & \textbf{CACC}$\uparrow$ & \textbf{PACC}$\uparrow$ & \textbf{ASR}$\downarrow$ \\ \midrule
\multirow{4}{*}{\textbf{SST-2}}  & \texttt{badnets}          & 86.16         & 81.14         & 18.86        \\
                                 & \texttt{addsent}          & 83.31         & 68.75         & 31.25        \\
                                 & stylebkd         & 83.80         & 46.27         & 53.73        \\
                                 & \texttt{synbkd}           & 81.88         & 57.24         & 42.76        \\ \midrule
\multirow{4}{*}{\textbf{YELP}}   & \texttt{badnets}          & 91.50         & 93.40         & 6.60         \\
                                 & \texttt{addsent}          & 91.40         & 80.75         & 19.25        \\
                                 & \texttt{stylebkd}         & 90.04         & 72.75         & 27.25        \\
                                 & \texttt{synbkd}           & 92.27         & 68.75         & 31.25        \\ \midrule
\multirow{4}{*}{\textbf{HSOL}}   & \texttt{badnets}          & 94.73         & 93.81         & 6.19         \\
                                 & \texttt{addsent}          & 95.17         & 90.43         & 9.57         \\
                                 & \texttt{stylebkd}         & 95.13         & 63.56         & 36.44        \\
                                 & \texttt{synbkd}           & 94.85         & 53.66         & 46.34        \\ \midrule
\multirow{4}{*}{\textbf{Agnews}} & \texttt{badnets}          & 91.93         & 90.18         & 4.02         \\
                                 & \texttt{addsent}          & 92.99         & 90.33         & 1.53         \\
                                 & \texttt{stylebkd}         & 93.03         & 80.89         & 10.16        \\
                                 & \texttt{synbkd}           & 93.09         & 72.54         & 11.81        \\ \bottomrule
\end{tabular}
\caption{Backdoor purification performance (in percentage) of our BadActs w/o the detection module. $\uparrow$ indicates higher is better and $\downarrow$ indicates lower is better.}
\vspace{-0.3cm}
\label{table: ablition_wo_detection}
\end{table*}